\documentclass{epl}

\newcommand{\btau}{\mbox{\boldmath{$\tau$}}}

\newcommand{\bh}{\mbox{\boldmath{$h$}}}

\newcommand{\bX}{\mbox{\boldmath{$X$}}}
\newcommand{\bD}{\mbox{\boldmath{$D$}}}
\newcommand{\bb}{\mbox{\boldmath{$b$}}}
\newcommand{\br}{\mbox{\boldmath{$r$}}}
\newcommand{\bbs}{\mbox{\boldmath{$s$}}}
\newcommand{\bS}{\mbox{\boldmath{$S$}}}

\newcommand{\bbm}{\mbox{\boldmath{$m$}}}

\newcommand{\bL}{\mbox{\boldmath{$L$}}}
\newcommand{\bA}{\mbox{\boldmath{$A$}}}
\newcommand{\ba}{\mbox{\boldmath{$a$}}}
\newcommand{\be}{\mbox{\boldmath{$e$}}}
\newcommand{\bR}{\mbox{\boldmath{$R$}}}
\newcommand{\bO}{\mbox{\boldmath{$O$}}}

\newcommand{\boeta}{\mbox{\boldmath{$\eta$}}}

\title{Analysis of CDMA systems that are characterized by eigenvalue
spectrum}
\shorttitle{Analysis of CDMA by eigenvalue spectrum}
\author{K. Takeda\inst{1} 
\and S. Uda\inst{2} \and Y. Kabashima\inst{1}}

\institute{                    
  \inst{1} Department of Computational Intelligence 
and Systems Science, 
Tokyo Institute of Technology - Yokohama 2268502, Japan\\
  \inst{2} The Institute of Japanese Union Scientists \& Enginners 
- 5-10-11 Sendagaya, Shibuyaku, Tokyo 1510051, Japan 
}

\begin{document}

\maketitle

\begin{abstract}
An approach to analyze the performance of the 
code division multiple access (CDMA) scheme,  
which is a core technology used in modern wireless 
communication systems, is provided. 
The approach characterizes the objective system
by the eigenvalue spectrum of a cross-correlation 
matrix composed of signature sequences used in CDMA communication, 
which enable us to handle a wider class of CDMA systems beyond 
the basic model reported by Tanaka
in \textit{Europhys. Lett.,}~\textbf{54}~(2001)~{540}. 
The utility of the scheme is shown by analyzing a system in which 
the generation of signature sequences 
is designed for enhancing the orthogonality. 
\end{abstract}

\section{Introduction}
Over the last decade, the scope of statistical mechanics
has rapidly expanded beyond its original goal 
of analyzing many-body problems that arise when dealing with material objects. 
Information theory is a major source of problems, and research activity aimed at solving these problems is becoming popular. 
Identifying information bits with Ising spins, 
many problems in information theory, such as error correcting/compression codes
\cite{Sourlas1989,KabaSaad1998,KabaSaad1999,NishimoriWong1999,MontanariSourlas2000,Montanari2000,Murayama2002,HosakaNishimoriKabashima2003,MurayamaOkada2003,Ciliberti2005}
and cryptosystems \cite{KabaMuraSaad2000,Kinzel2003}, 
can be formulated as virtual 
many-body systems that are subject to disordered interactions. 
Analysis of the formulated problems using 
techniques of statistical mechanics
has provided various nontrivial results 
that have not been obtained by conventional methods 
of information theory \cite{Nishimori2001,KabaSaad2004}. 

Code division multiple access (CDMA), which is a core technology 
used in modern wireless communication, is an example of the successful application of such a statistical mechanical approach. This technology realizes simultaneous communication 
between multiple users and a single 
base station by modulating each user's bit signal (symbol)
into a sequence of random pattern, termed 
the signature sequence \cite{Verdu1998}. 
CDMA has already been employed in third-generation 
mobile phone systems and wireless LANs.

Tanaka (2001) showed that the replica 
method of statistical mechanics enables 
the accurate assessment of the communication performance 
of a basic 
CDMA model in which users' sequences
 are generated independently of 
each other in a large system limit \cite{Tanaka2001,Tanaka2002}. 
This research was considered to be ground-breaking, 
and variations of the basic model are currently being actively 
analyzed based on the scheme given by Tanaka 
\cite{GuoVerdu2005,Wenetal2005,Wen2006,Guo2006}. However, Tanaka's 
scheme relies on the assumption of statistical 
independence among users' sequences and, therefore,  
cannot be applied to systems in which sequences are not 
independent. As certain dependence generally 
arises among sequences when the system is designed 
for optimizing communication performance, 
this limitation is problematic with respect to practical relevance. 

The purpose of this Letter is to resolve this problem.
More precisely, we present another approach by which to analytically 
assess the performance of CDMA communication. The central concept of 
our scheme is to characterize the objective 
system by the eigenvalue spectrum of the cross-correlation 
matrix composed of sequences under the assumption 
that the orthogonal eigenbasis is randomly provided. 
The range of applicability of such characterization is
not clear at the current moment; however, this enables us to 
analytically handle CDMA systems regardless of whether sequences 
are statistically independent and offers, at least, a useful 
approximation scheme for investigating advanced systems. 
The utility of the approach is shown by analyzing 
a system that is difficult to assess accurately 
by the conventional scheme.

\section{Modeling}
As a general scenario of CDMA systems, let us assume
a situation in which $K$ users simultaneously transmit 
symbols $b_k^0=\pm 1$ $(k=1,2,\ldots,K)$ to 
a single base station utilizing the sequences of $N$-dimensional
vectors $\bbs_k=(s_{\mu k})$ $(\mu=1,2,\ldots,N)$, where 
$\mu$ indexes the chip timing of the symbol modulation
and normalization constraint $|\bbs_k|=1$ is imposed.  
We denote an $N \times K$ matrix composed of $\bbs_k$ 
and a set of the received amplitudes of 
users' signals as $\bS=(\bbs_k)$ and  a diagonal matrix $\bA={\rm diag}(A_k)$, 
respectively, both of which are assumed to be known to the base station. 
In order to cope with cases in which 
the sequences correlate with each other, 
we characterize the system by 
a $K \times K$ cross-correlation matrix 
$\bR=\bA^T \bS^T \bS \bA$, where $\bX^T$ represents the transpose
of $\bX$, and \textit{assume} that 
$\bR$ can be regarded as 
\begin{eqnarray}
\bR=\bO   \bD  \bO^{-1}, 
\label{random_orthogonal}
\end{eqnarray}
where $\bD$ is a diagonal matrix $\bD={\rm diag }(D_k)$, 
the eigenvalue spectrum of which, $K^{-1}\sum_{k=1}^K\delta (\lambda-D_k)$,
is provided as $\rho(\lambda)$ and a $K \times K$ orthogonal matrix $\bO$ is 
assumed to be drawn randomly from the uniform distribution 
defined by the Haar measure on the orthogonal 
group \cite{ParisiPotters1995}. 
Such characterization is supposed to be plausible because 
for a wide class of systems $\bR$ becomes dense and 
the sequences are generated in a somewhat random manner, 
which statistically produces no preferential direction, 
although mathematical validity of assuming Eq. (\ref{random_orthogonal}) 
should be examined in future research.
Under the additional simplifying assumptions that the channel noise 
is the additive white Gaussian of variance 
$\sigma_0^2$ and both the chip timing and symbol timing 
are synchronous across all users, 
the received signals at base station, $\br=(r_\mu)$, 
can be represented as 
\begin{eqnarray}
\br=\bS \bA \bb^0 +\sigma_0 \boeta, 
\label{received_signal}
\end{eqnarray}
where $\bb^0=(b_k^0)$ and $\boeta$ is an $N$-dimensional 
random vector, each component of which is generated 
independently from the normal distribution ${\cal N}(0,1)$. 

After receiving $\br$, the remaining task of the base station is 
to infer the users' original symbols $\bb^0$, 
which is called user detection or demodulation. 
The optimal demodulation scheme to minimize the bit-wise 
probability of incorrect estimation, which is referred to as the 
bit error rate $P_b$, is provided by using 
the posterior distribution 
\begin{eqnarray}
P(\bb|\br)=Z^{-1}
\exp \left [
-\frac{1}{2\sigma^2}
(\br-\bS \bA \bb)^T (\br-\bS \bA \bb) \right ], 
\label{posterior}
\end{eqnarray}
as $\hat{b}_k={\rm sign}\left (\sum_{\bb   } b_k
P(\bb|\br) \right )$
when the assumed 
noise variance $\sigma^2$ is in accordance with the correct value 
$\sigma^2_0$. 
Here, $Z=\sum_{\bb}\exp \left [
-(\br-\bS \bA \bb)^T (\br-\bS \bA \bb)
/(2\sigma^2) \right ]$ 
serves as the partition function, 
$\hat{b}_k$ denotes the estimate of the $k$th user's symbol
$b_k^0$ and ${\rm sign}(u)=u/|u|$ for $u \ne 0$. 

\section{Analysis}
Since Eq. (\ref{posterior}) is conditioned by 
the predetermined random variable 
$\br=\bS \bA \bb^0 +\sigma_0 \boeta$, we resort to 
the replica method to assess the typical performance of the CDMA system
assuming a large system limit $K,N \to \infty$ while maintaining 
the load $\beta= K/N$. Namely, we 
evaluate the $n(=1,2,\ldots)$th moment of the partition 
function $Z$ with respect to the relevant 
predetermined randomness and 
analytically continue the obtained expression to real 
$n \in {\bf R}$. For this purpose, it is convenient to first take 
the average with respect to $\boeta$, which provides
\begin{eqnarray}
\int D\boeta \exp \left [-\frac{1}{2\sigma^2}
\sum_{a=1}^n (\br-\bS \bA \bb^a)^T (\br-\bS \bA \bb^a) \right ]
=\exp \left [ \frac{1}{2} {\rm Tr} \bR \bL(n) \right ], 
\label{Ln}
\end{eqnarray}
where $\bb^a$ is the $a$th replica of symbol vector $\bb$, 
$D\boeta=(2 \pi )^{-N/2}\exp \left [-|\boeta|^2/2 \right ]$
and 
\begin{eqnarray}
\bL(n)=-\frac{1}{\sigma^2}
\sum_{a=1}^n (\bb^0-\bb^a)(\bb^0-\bb^a)^T
+\frac{\sigma^2_0}{\sigma^2(\sigma^2+n \sigma_0^2)}
\left (\sum_{a=1}^n (\bb^0-\bb^a) \right )
\left (\sum_{b=1}^n (\bb^0-\bb^b) \right )^T. 
\label{operator}
\end{eqnarray}
Equation (\ref{random_orthogonal}) indicates that the average of 
Eq. (\ref{Ln}) with respect to the cross-correlation matrix
$\bR$ is reduced to a form of 
\begin{eqnarray}
\int {\cal D}\bR \exp \left [ \frac{1}{2} {\rm Tr} \bR \bL(n) 
\right ]\simeq \exp \left [ K {\rm Tr} G\left 
(\frac{\bL(n)}{K} \right )
\right ], 
\label{Gexpresssion}
\end{eqnarray}
for large $K$, where a function $G(x)$ is provided 
by the eigenvalue spectrum $\rho (\lambda)$ as
$G(x)=(1/2)\int_0^x dt \left (\Lambda (t)-t^{-1}
\right )$, 
utilizing a function $\Lambda (x)$ that is implicitly 
determined by the condition \cite{ParisiPotters1995,OpperWinther2001}
\begin{eqnarray}
x=\int d\lambda \rho(\lambda)
(\Lambda(x)-\lambda)^{-1}. 
\label{Lambdax}
\end{eqnarray}
For a given $G$-function, $G(x)$, Eq. (\ref{Lambdax}) 
(Cauchy transform) can also be utilized to assess $x$
 as a function of $\Lambda$. 
The function $x(\Lambda)$ is directly linked to the spectrum
by Stieltjes inversion formula,
\begin{eqnarray}
\rho(\lambda)=\lim_{\epsilon \to +0} 
\frac{1}{\pi}{\rm Im}\ x(\lambda-\sqrt{-1} \epsilon). 
\label{x_to_rho}
\end{eqnarray}

Intrinsic permutation symmetry among replicas 
naturally leads to the replica symmetric
(RS) ansatz, implying that 
configurations characterized 
by $\bb^0 \cdot \bb^a=K m$ $(a=1,2,\ldots,n)$ 
and $\bb^a \cdot \bb^b =K q$ $(a>b)$
provide the most dominant contribution to the moment evaluation. 
Under this ansatz, the $K \times K$ matrix $\bL (n)$ 
has three types
of eigenvalues: $\lambda_1=-K(\sigma^2+n \sigma_o^2)^{-1}
(1-q+n(1-2m+q))$, 
$\lambda_2=-K\sigma^{-2}(1-q)$ and $\lambda_3=0$, 
the numbers of degeneracy of which are 
$1$, $n-1$ and $K-n$, respectively. 
This indicates that Eq. (\ref{Gexpresssion}) is evaluated as
\begin{eqnarray}
\exp \left [ 
K \left (
G \left (-\frac{1-q+n(1-2m+q)}{\sigma^2+n \sigma_o^2} \right )
+(n-1) G\left (-\frac{1-q}{\sigma^2} \right ) \right ) \right ]. 
\label{eigenvalues}
\end{eqnarray}
Continuing both of this and 
the RS entropic contribution of 
${\rm Tr}_{\bb^1,\bb^2,\ldots,\bb^n}
\prod_{a=1}^n \delta(\bb^0\cdot \bb^a-Km)
\prod_{a>b} \delta (\bb^a \cdot \bb^b -K q)$
analytically from $n=1,2,\ldots$ to real values, 
$n \in {\bf R}$, provides an expression of
the average free energy
\begin{eqnarray}
\frac{1}{K} \overline{\ln Z} &=&\lim_{n \to 0}
\frac{1}{nK} \ln \overline{Z^n} \cr 
&=&\mathop{\rm Extr}_{m,q,\hat{m},\hat{q}}
\left \{
G\left (-\frac{1-q}{\sigma^2} \right )
+\left (-\frac{1-2m+q}{\sigma^2}+\frac{\sigma_0^2(1-q)}{\sigma^4}
\right )
G^\prime \left (-\frac{1-q}{\sigma^2} \right ) \right . \cr
&& \phantom{aaaaaaaaaaaaa}
\left . -\hat{m}m-\frac{\hat{q}(1-q)}{2}
+\int Dz \ln \left (2 \cosh \sqrt{\hat{q}} z+\hat{m} \right )
\right \}, 
\label{free_energy}
\end{eqnarray}
where $Dz=(2\pi)^{-1/2} dz \exp \left [ -z^2/2 \right ]$,
$\mathop{\rm Extr}_{u} \{ \cdots  \}$ indicates 
the extremization of $\{ \cdots \}$ with respect to $u$, 
and $\overline{(\cdots)}$ denotes 
the average of $(\cdots)$ with respect to $\boeta$ and $\bR$. 
Equation (\ref{free_energy}) is the main result of the present letter. 

Three things are noteworthy here. First, the bit error rate 
$P_b$ can be assessed as 
$P_b=\int_{\hat{m}/\sqrt{\hat{q}}}^{\infty} Dz $ 
by utilizing the values of the conjugate 
variables $\hat{q}$ and $\hat{m}$ that extremize
Eq. (\ref{free_energy}). 
This implies that Eqs. (\ref{Lambdax}) and (\ref{free_energy})
offer a complete scheme for assessing 
the typical communication performance of CDMA systems
characterized by the eigenvalue spectrum $\rho(\lambda)$. 
Second, the basic model analyzed in \cite{Tanaka2001} 
can be characterized 
by $\rho(\lambda)=\rho_{\rm basic}(\lambda)=
[1-\beta^{-1}]^+\delta(\lambda)+(2 \pi \beta \lambda)^{-1}\sqrt{
\left [\lambda-(1-\sqrt{\beta})^2\right ]^+
\left [(1+\sqrt{\beta})^2-\lambda \right ]^+}$, where
$[ u ]^+= u$ for $u > 0$ and $0$, otherwise. 
This yields $G(x)=G_{\rm basic}(x)=
-(2\beta)^{-1}\ln (1-\beta x)$, which, when inserted into 
Eq. (\ref{free_energy}), reproduces an expression 
equivalent to the free energy of \cite{Tanaka2001} 
(Eq. (10) in \cite{Tanaka2001}). 
Namely, the conventional analysis is one component of our formalism. 
Finally, although thus far we have assumed the RS ansatz, 
the local stability of the RS solution can be broken 
when the assumed noise variance $\sigma^2$ is set 
sufficiently smaller than the correct value $\sigma_0^2$. 
The stability condition with respect to the local perturbation 
of breaking the replica symmetry can be examined via 
the de Almeida-Thouless analysis \cite{AT1978}, 
which, in the current case, yields the following expression: 
\begin{eqnarray}
1-\frac{2}{\sigma^4}G^{\prime \prime}
\left (-\frac{1-q}{\sigma^2} \right )
\int Dz \left (1-\tanh^2(\sqrt{\hat{q}}z+\hat{m})\right )^2 >0 . 
\label{ATline}
\end{eqnarray}
A similar result was reported in \cite{OpperWinther2001}. 
\section{Computationally feasible demodulation algorithm} 
Although the above mentioned procedure
provides the typical performance of the estimator 
$\hat{b}_k={\rm sign}\left (\sum_{\bb   } b_k
P(\bb|\br) \right )$, performing the procedure exactly for a given 
received signal $\br$ is computationally difficult 
in the general case. Advanced mean field methods of the 
Thouless-Anderson-Palmer (TAP) type have recently been 
receiving attention as promising approaches by which to resolve this 
difficulty \cite{TAP1977,OpperSaad2001}. 
For the basic models, one can 
construct a computationally feasible demodulation 
algorithm along this line by using 
the property of statistical independence of sequences\cite{Kabashima2003}. 
Unfortunately, this strategy is not 
available in the present case. Nevertheless, 
one can construct such a heuristic algorithm that 
coincides with the result for the basic model 
when $\rho(\lambda)$ is set to $\rho_{\rm basic}(\lambda)$. 
Denoting the approximate symbol average 
$\sum_{\bb   } b_k P(\bb|\br)$ 
at the $t$th update as $m_k^t=\tanh(h_k^t)$, 
we obtain 
\begin{eqnarray}
\begin{array}{rcl}
\ba^{t+1}&=&2 G^\prime \left (-\frac{1-q^{t}}{\sigma^2}
\right )\left [
\frac{1}{\sigma^2}\left (\br - \bS \bA \bbm^t \right )+
\left (\frac{1}{2 G^\prime \left (-\frac{1-q^{t}}{\sigma^2}
\right )}-1 \right )
 \ba^t \right ], \cr
\bh^t&=&\bA^T \bS^T \ba^t+\frac{2}{\sigma^2}G^\prime 
\left (-\frac{1-q^{t-1}}{\sigma^2} 
\right ) \bbm^{t-1}, 
\end{array}
\label{TAP}
\end{eqnarray}
where $\bbm^t=(m_k^t)$, $\bh^t=(h_k^t)$, $\ba^t=(a_\mu^t)$
and $q^t=K^{-1}\sum_{k=1}^K (m_k^t)^2$. 
The estimate of $\hat{b}_k$ at the 
$t$th update is provided as $\hat{b}_k={\rm sign}(h_k^t)$. 
Numerical experiments indicate that Eq. (\ref{TAP}) 
empirically converges by $O(1)$ updates 
in most cases of large systems 
when the ratio $\sigma_0^2/\sigma^2$ 
is sufficiently low. 
This implies that one can approximately perform 
the demodulation by serial computers in a time scale of $O(NK)$. Note that Eq. (\ref{TAP}) is simply an example of the dynamics,
the fixed point of which agrees with the solution 
of the TAP equation in general, whereas 
several favorable properties hold 
in the specific case of $\rho(\lambda)=\rho_{\rm basic}(\lambda)$
\cite{Tanaka2005}. 
The quest for better algorithms of this sort is currently 
under way, as will be reported elsewhere. 

\section{Example} In order to address the significance of the developed
approach, we consider a CDMA system in which 
the generation of sequences is devised for the purpose of enhancing the orthogonality. 
In this system, each $\bbs_k$ is constructed so as to be 
orthogonal to all of the existing sequences 
$\bbs_1,\bbs_2, \ldots, \bbs_{k-1}$ until $k$ reaches $N$. 
Sequence generation of this sort becomes impossible 
when $k$ exceeds $N$.
Therefore, $\bbs_{N+1}$ is randomly 
generated and $\bbs_k$ for $N+2 \le k \le 2N$ is 
constructed so as to be orthogonal 
to $\bbs_{N+1}, \bbs_{N+2}, \ldots, \bbs_{k-1}$
without regard to the sequences of the previous group, 
$\bbs_1,\bbs_2, \ldots, \bbs_{N}$. 
For $k \ge 2N+1$, this procedure is repeated. 
Sequence generation of this sort was once considered with respect to
the improvement of the learning performance of 
linear perceptrons \cite{Sollich1994}. 
For $N=2^L$ ($L=1,2,\ldots)$, 
such generation is easily implemented by
drawing a random gauge vector 
$\btau_g=(\tau_{\mu g})=
\{+1,-1\}^N$ for each group $g(=1,2,\ldots)$ of 
size $N$ and assigning the sequences of group 
$g=\lfloor k/N \rfloor+1$ as 
$\bbs_{k}=(\tau_{\mu \lfloor k/N \rfloor} 
e_{\mu \Delta(k,N)})$
by utilizing the normalized Walsh-Hadamard basis 
$\be_1,\be_2,\ldots,\be_N$ \cite{Harmuth1970}. 
Here, $\lfloor u \rfloor$ indicates the maximum integer that 
does not exceed $u$, and $\Delta(k,N)=k-N \lfloor k/N \rfloor$. 
The orthogonality 
perfectly cancels the crosstalk 
noise from other users of the identical group, which is advantageous for the improvement of communication performance. However, the orthogonality constraint yields strong dependence among sequences of an identical group, 
which prevents the 
scheme provided in \cite{Tanaka2001} from accurately 
assessing the performance. 

\begin{figure}[t]
\setlength{\unitlength}{1mm}
\begin{picture}(180,39)
\put(15,38){\includegraphics[width=40mm,angle=270]{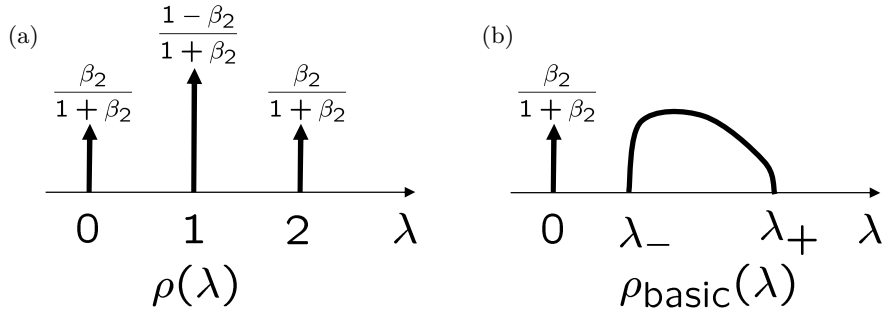}}
\put(10,33){(a)}
\put(73,33){(b)}
\end{picture}
\caption{Schematic diagrams of eigenvalue spectra for (a) the current
 example system of $\beta_1=1$, $\beta_2 =\beta - 1\ge 0$ and $\beta_k
 =0$ $\forall{k}\ge 3$ and (b) the basic model of \cite{Tanaka2001}, 
 in which the component of the sequences, $s_{\mu k}$, is provided as
 $+ N^{-1/2}$ or $-N^{-1/2}$ with an equal probability independently of 
 a pair of indices $\mu, k$. Amplitudes $A_k$ are set to unity $\forall{k}$
 in both systems. Even in the basic model, a randomly drawn pair of
 sequences,  $\bbs_i$ and $\bbs_j$, is regarded as {\em almost}
 orthogonal for large $N$ since the overlap $\bbs_i\cdot \bbs_k$ typically 
 vanishes as $O(N^{-1/2})$ as $N \to \infty$. 
 However, the constraint of perfect orthogonality within an identical
 group that is imposed in the example system makes the feature of the
 eigenvalue spectrum $\rho (\lambda)$, which is composed of three delta-peaks
 at $\lambda =0$, $1$ and $2$, considerably different from that of
 $\rho_{\rm basic}(\lambda)$, which is made up of a single delta-peak at
 $\lambda=0$ and a continuous part between
 $\lambda_{+}=(1+\sqrt{\beta})^2$ and $\lambda_{-}=(1-\sqrt{\beta})^2$. }
\label{fig1}
\end{figure}

On the other hand, our approach potentiates 
performance analysis by evaluating the eigenvalue 
spectrum $\rho(\lambda)$ of $\bR$ 
under the assumption of Eq. (\ref{random_orthogonal}), 
which is expected to hold since the sequence generation 
statistically creates no directional preference. 
For this purpose, we fist evaluate another spectrum 
$\tilde{\rho}(\lambda)$ of $\tilde{\bR}=\bS \bA\bA^T\bS^T$, 
which is dual to $\bR$ and easier to handle. 
In CDMA systems, a set of users that communicate with 
a given base station generally varies in time. 
Let us denote the set of user indices
that remain in group $g$ as ${\cal K}(g)$ at the objective 
moment. We also introduce another notation 
$\beta_g=|{\cal K}(g)|/N$, which implies that 
$0 \le \beta_g \le 1$ $\forall{g}$ and 
$\beta=\sum_g \beta_g$. Using these, the dual matrix can be 
expressed as $\tilde{\bR}=\sum_g \tilde{\bR}_g$, where 
$\tilde{\bR}_g=\sum_{k \in {\cal K}(g)}A_k^2 \bbs_k 
\bbs_k^T$, the eigenvalue spectrum of which 
is provided as $\tilde{\rho}_g(\lambda)=(1-\beta_g)\delta(\lambda)
+N^{-1}\sum_{k \in {\cal K}(g)}\delta(\lambda-A_k^2)$. 
We express the $G$-function of $\tilde{\bR}_g$, 
which is given by $\tilde{\rho}_g 
(\lambda)$ via Eq. (\ref{Lambdax}), as
$\tilde{G}_g(x)$. Statistical independence across the groups
guarantees that the $G$-function of $\tilde{\bR}$, $\tilde{G}(x)$, 
is assessed as $\tilde{G}(x)=\sum_g \tilde{G}_g(x)$. 
This provides the spectrum $\tilde{\rho}(\lambda)$ 
through Eq. (\ref{x_to_rho}), which yields 
the spectrum of ${\bR}$ as
$\rho(\lambda)=[1-\beta^{-1}]^+\delta(\lambda)+\beta^{-1}
\tilde{\rho}(\lambda)$. 

\begin{figure}[t]
\setlength{\unitlength}{1mm}
\begin{picture}(180,40)
\put(0,46){\includegraphics[width=45mm,angle=270]{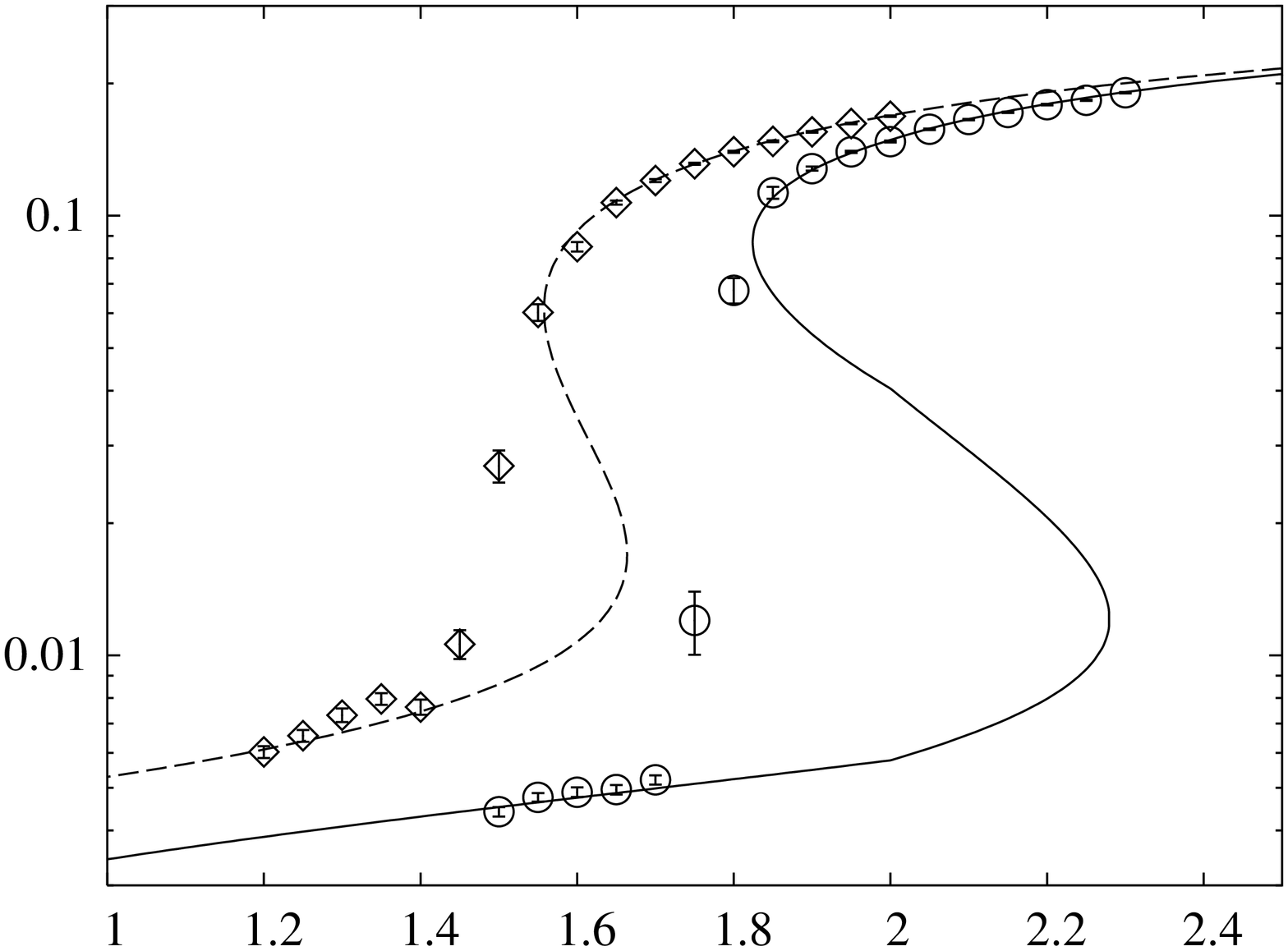}}
\put(2,24){$P_b$}
\put(27,-2){$\beta=K/N$}
\put(1,41){(a)}
\put(73,45){\includegraphics[width=42mm,angle=270]{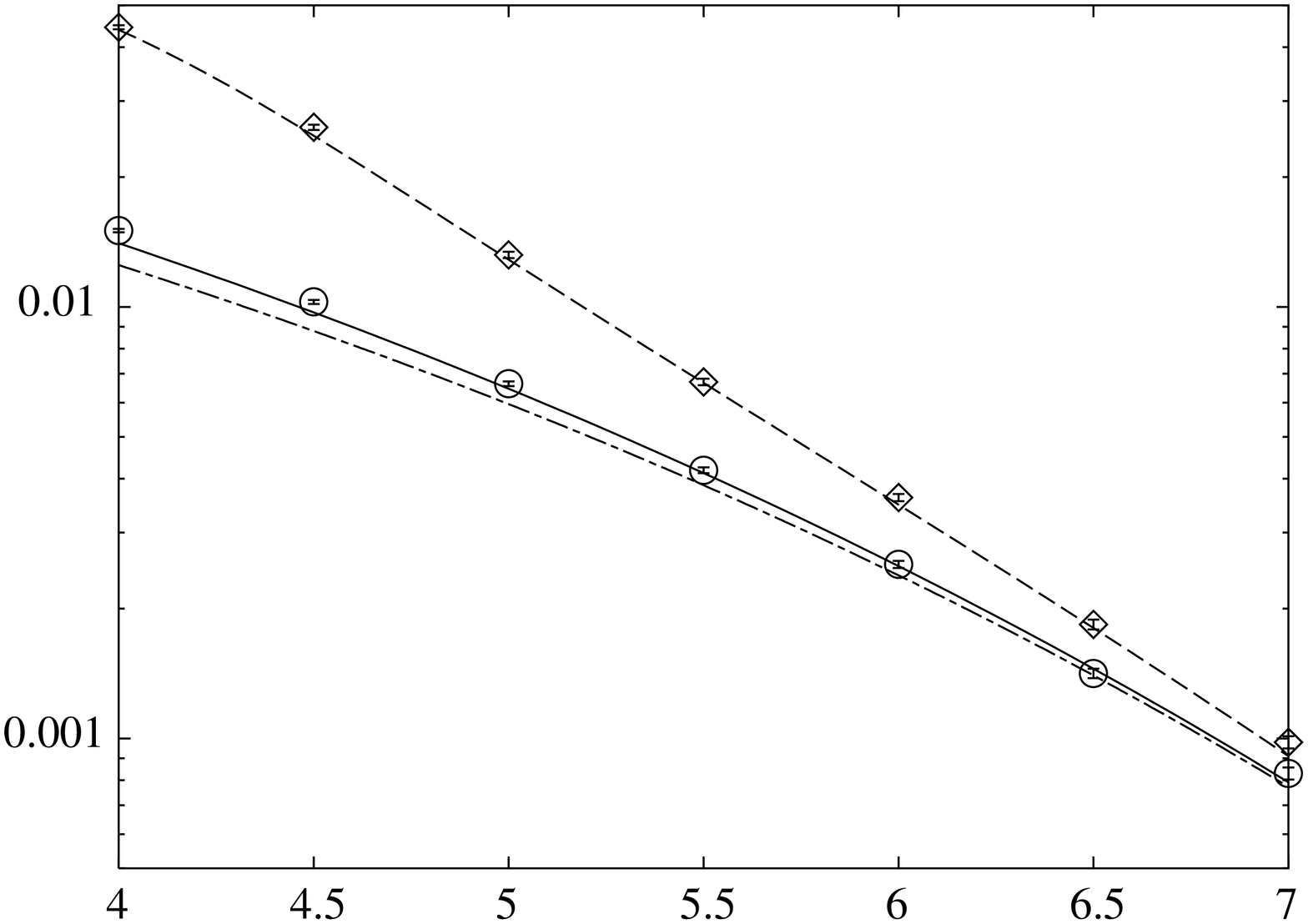}}
\put(73,24){$P_b$}
\put(100,-2){$E_b/N_0$ [dB]}
\put(71,41){(b)}
\end{picture}
\caption{(a) Bit error rate $P_b$ vs. the load $\beta=K/N$
under the condition of $\sigma_0=\sigma=0.37$. 
The solid and broken lines indicate the theoretical prediction
for the example system and the basic model, respectively.
Because of the reduction of the crosstalk noise,
a lower bit error rate can be achieved in the example system 
than that of the basic model for any load $\beta$.
In addition, the example system provides
a larger value of the spinodal point
than that of the basic model, which implies that 
the example system can deal with more users by 
computationally feasible algorithms than 
the basic model. The circles and diamonds represent 
the results obtained from 200 experiments 
utilizing variations of Eq. (\ref{TAP})
for the example system and basic model of $N=2048$, respectively, 
in which initial states were set to the matched 
filter output $h_k^0=\bbs_k \cdot \br$ following 
the conventional scheme \cite{Kabashima2003,Tanaka2005}. 
These exhibit excellent agreement with theoretical 
curves, except around the spinodal points, where 
large fluctuations make it difficult for the algorithm to converge. 
(b) Bit error rate $P_b$ vs. $E_b/N_0=10 
\log_{10}[1/(2\sigma^2)]$, which is a conventional measure 
of the signal-to-noise ratio, for $\beta=1.1$. Starting from the bottom
 of the figure, the three curves indicate the achievable limit
(performance when the channel is accessed by only a single user), 
the theoretical predictions 
of the example system and the basic model, 
respectively. The circles and diamonds represent 
the numerical results obtained from 500 experiments 
on the basis of Eq. (\ref{TAP}) for the example system, 
which exhibits excellent consistency with the theoretical
prediction. 
}
\label{fig2}
\end{figure}

As a simple but nontrivial example, we examined the 
case of $\beta_g=1$, $\beta-\lfloor \beta \rfloor$ and
$0$ for $g(\ge 0)$ of being less than, equal to and 
greater than  $\lfloor \beta \rfloor +1$, respectively, 
and $A_k=1$ $\forall{k}$, 
which means $\rho(\lambda)=\left [
1-\beta^{-1} \right ]^+
\delta(\lambda) + 
\beta^{-1}\left ((1-\beta+\lfloor \beta \rfloor)
\delta (\lambda-\lfloor \beta \rfloor)
+(\beta-\lfloor \beta \rfloor)
\delta(\lambda-\lfloor \beta \rfloor-1)
\right )$ (Fig. \ref{fig1}). 
Figure \ref{fig2} shows that the current system of 
the random orthogonal sequence generation 
outperforms the basic model 
with respect to both the bit error rate and the position of 
the spinodal point, which practically determines 
the limitation of computationally feasible demodulation. 
Numerical experiments 
based on variations of Eq. (\ref{TAP}), 
the results of which are plotted as symbols, 
exhibit excellent agreement with the theoretical 
predictions, which are denoted by the curves, 
indicating the significance of our approach. 
 
\section{Summary}
In summary, we have presented a novel approach by which to analyze 
the performance of CDMA communication systems. 
In the provided scheme, the properties of CDMA systems are
characterized by the eigenvalue spectrum of the 
cross-correlation matrix composed of signature sequences. 
This enables us to accurately analyze a wide class of CDMA 
systems, even when signature sequences are dependent on each other. 
The significance of our approach has been demonstrated by application to a system that cannot be handled by the conventional scheme, 
the results of which showed excellent agreement with numerical experiments. 

Application to various cases and generalization to 
asynchronous communication will be examined in future studies. 
\acknowledgments
This study was supported in part by Grant-in-Aids
Nos. 17340116 and 18079006 from JSPS/MEXT, Japan (YK).

\end{document}